\documentclass[a4paper]{article}

\usepackage{INTERSPEECH2019}
\usepackage[utf8]{inputenc}
\usepackage{todonotes}
\usepackage{textcomp}

\title{Avaya Conversational Intelligence\texttrademark: A Real-Time System for Spoken Language Understanding in Human-Human Call Center Conversations}

\name{Jan Mizgajski$^1$$^3$, Adrian Szymczak$^1$, Robert Głowski$^1$, Piotr Szymański$^1$$^2$, Piotr Żelasko$^1$$^4$, Łukasz Augustyniak$^1$$^2$, Mikołaj Morzy$^1$$^3$, Yishay Carmiel$^1$, Jeff Hodson$^1$, Łukasz Wójciak$^1$, Daniel Smoczyk$^1$, Adam Wróbel$^1$, Bartosz Borowik$^1$, Adam Artajew$^1$, Marcin Baran$^1$, Cezary Kwiatkowski$^1$, Marzena Żyła-Hoppe$^1$}
\address{
  $^1$Avaya, Santa Clara, California, United States\\
  $^2$Wrocław University of Science and Technology, Wrocław, Poland\\
  $^3$Poznan University of Technology, Poznań, Poland\\
  $^4$AGH University of Science and Technology, Kraków, Poland}
\email{jmizgajski@avaya.com}

\begin{document}

\maketitle


\begin{abstract}
Avaya Conversational Intelligence\texttrademark~(ACI) is an end-to-end, cloud-based solution for real-time Spoken Language Understanding for call centers. It combines large vocabulary, real-time speech recognition, transcript refinement, and entity and intent recognition in order to convert live audio into a rich, actionable stream of structured events. These events can be further leveraged with a business rules engine, thus serving as a foundation for real-time supervision and assistance applications. After the ingestion, calls are enriched with unsupervised keyword extraction, abstractive summarization, and business-defined attributes, enabling offline use cases, such as business intelligence, topic mining, full-text search, quality assurance, and agent training. ACI comes with a pretrained, configurable library of hundreds of intents and a robust intent training environment that allows for efficient, cost-effective creation and customization of customer-specific intents.
\end{abstract}


\section{Introduction}

Call center conversations are a valuable, but still underutilized asset for organizations.
The majority of speech recognition products focus on offline use cases, such as speech analytics, quality assurance, or agent training. This means that business process changes, potentially identified during the analysis, have to be enforced outside of the application. With real-time conversation understanding, organizations can recognize the problem and act during the call, enabling such use cases as real-time supervision and coaching, agent assistants and robotic process automation, or live compliance and brand risk monitoring. We present an end-to-end system for real-time call center conversation understanding and several use cases built on top of it. 

\section{Overview of Features and Capabilities}

\subsection{Real-Time Capabilities}

Avaya Conversational Intelligence\texttrademark~has the following real-time capabilities.

\textbf{Large Vocabulary Speech Recognition}, which is based on Kaldi~\cite{Povey:192584} and our proprietary models, is capable of transcribing thousands of concurrent calls with very low phrase latency, providing high accuracy and computing efficiency.

\textbf{Transcript Refinement} - the transcript is further refined with automatic: punctuation (prediction of question marks, periods, and commas) \cite{DBLP:conf/interspeech/ZelaskoSMSCD18}, truecasing (recovering the proper casing of recognized words), and readability turns (a word sorting algorithm to improve the readability of overlapping utterances from multiple speakers).

\textbf{Intent and Entity Recognition} - our Spoken Language Understanding engine comes with a fast and easy to train intent and entity recognition accompanied by a~library of hundreds of pretrained intents. With capabilities such as: fuzzy matching, support for long intents (7+ words), recovery from speech recognition errors, and entity parsing (matching entities as parts of intents and parsing them into a machine-readable format), it is able to capture complex intents with high precision and recall. Entities are divided into two robust categories: \textit{system entities} (numbers, amounts of money, dates, duration, quantities, volumes, percentages, spelling, names, surnames, locations) and \textit{user-defined entities} – placeholders, such as BRAND or PRODUCT, that can be filled in by the customer, thus lowering the barrier of entry in both customizing existing intents and defining new ones.

\textbf{Business Rules Engine} - business rules model sequences and aggregates over the stream of events, such as intents, entities, or call events, to allow organizations to define complex triggers used to power such use cases as real-time supervision or agent assistance.

\subsection{Post-Call Capabilities}

The real-time capabilities are complemented with post-call analysis.

\textbf{Unsupervised Keyphrase Extraction} - we run a proprietary unsupervised keyphrase extraction algorithm to extract additional context for analytics to better service out-of-ordinary cases that are not covered by the configured set of expected intents.

\textbf{Call Summary} - calls are summarized using natural language generation based on all extracted information, such as intents, entities, and keyphrases. 

\subsection{Batch Capabilities}

For customers that wish to work with recordings rather than live audio, we offer speaker diarization and classification (customer, call center agent), as well as all real-time and post-call capabilities in batch processing mode.

\subsection{Aggregation, Search, and Business Intelligence}

Based on the data extracted from multiple conversations, we offer:

\begin{itemize}
    \item information retrieval with aggregation and filtering on multiple criteria, such as full-text search, matched intents and entities, call metadata (agents, projects, etc.), customer-supplied metrics and labels
    \item hierarchical topic mining
    \item visualizations and dashboards
\end{itemize}

\subsection{Built-In Intent Training System}

To allow fast training and customization, our product comes with built-in training including:
\begin{itemize}
    \item mass call annotation with phrase grammars
    \item automatic, self-adapting synonym recommendations
    \item support for entities as parts of phrases
    \item instant verification on historical data sets
    \item easy-to-use UI, optimized for rapid intent definition
    \item feedback loops and false-positive training
\end{itemize}
\section{Key Applications}

In this section we present some of the key applications of Avaya Conversational Intelligence\texttrademark.

\subsection{Sentinel}

Sentinel is a dashboard of live calls showing the most recent intents. Calls are assigned risk scores based on customer-defined business rules that can be triggered by various events in the call (intents, entities, etc.). High-risk calls are highlighted and enriched by additional context. A supervisor can start watching the transcription and events of the problematic call and take corrective action.  

\begin{figure}[!ht]
\centering
\includegraphics[scale=0.2]{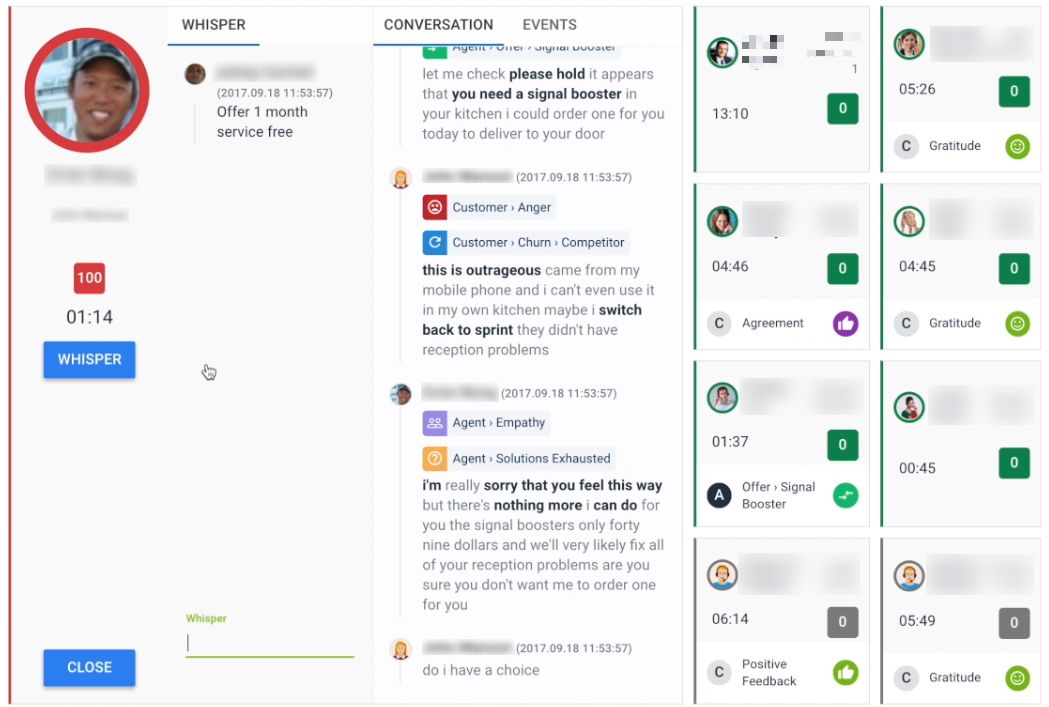}
\caption{Sentinel application, showing live supervision of a~problematic call.}
\label{fig:sentinel}
\end{figure}

\subsection{Explorer}

With Explorer, our customers can query, aggregate, filter, and visualize all historical calls. Each query can be saved and turned into a dashboard widget for future reference or comparison with multiple queries. If a particular call proves to be interesting, the customer can inspect it in detail along with its recording, transcription, and all extracted data.

\begin{figure}[!ht]
\centering
\includegraphics[scale=0.15]{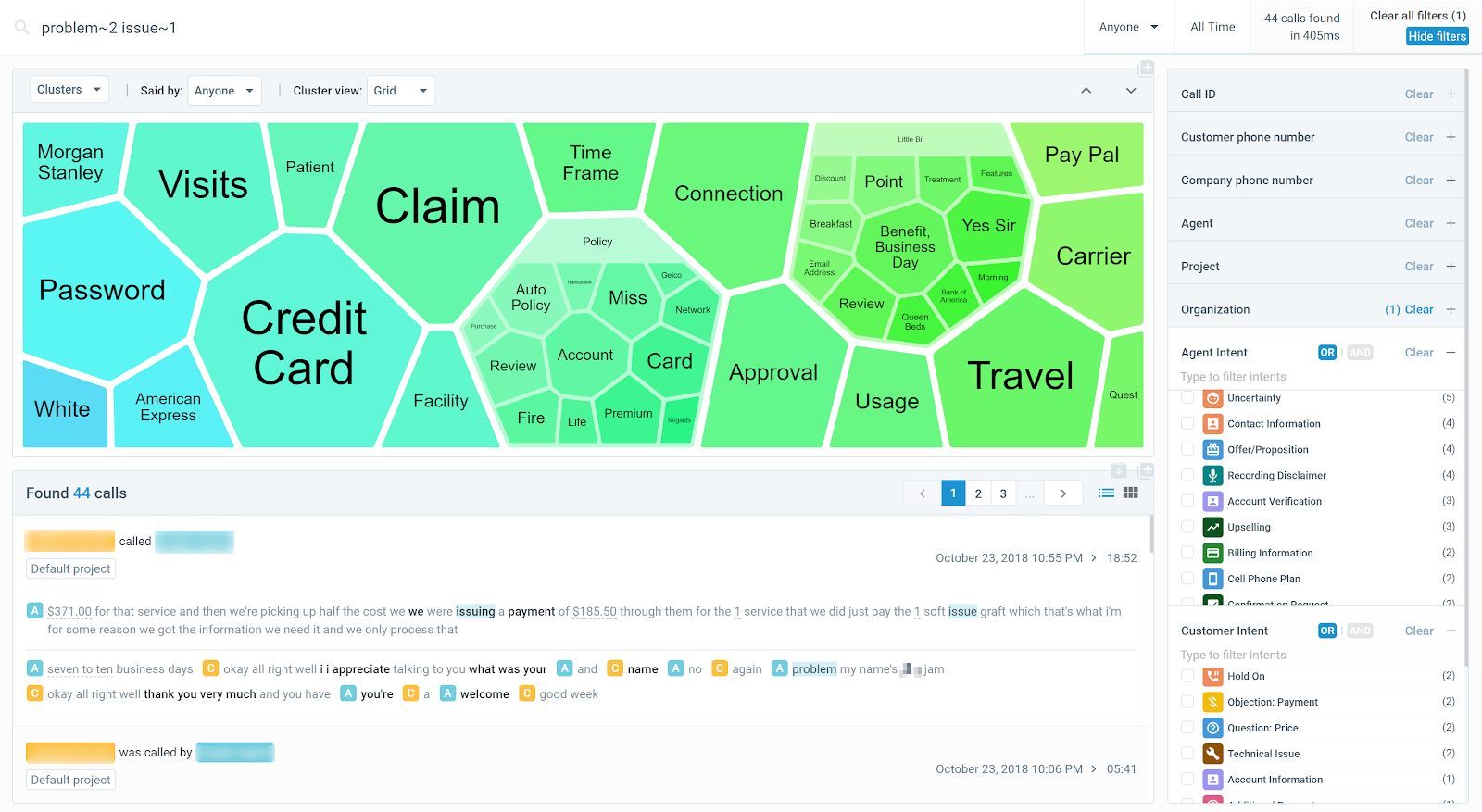}
\caption{The Explorer app showing results for a query with topic mining visualization and some available filters.}
\label{fig:explorer}
\end{figure}

\begin{figure}[!ht]
\centering
\includegraphics[scale=0.15]{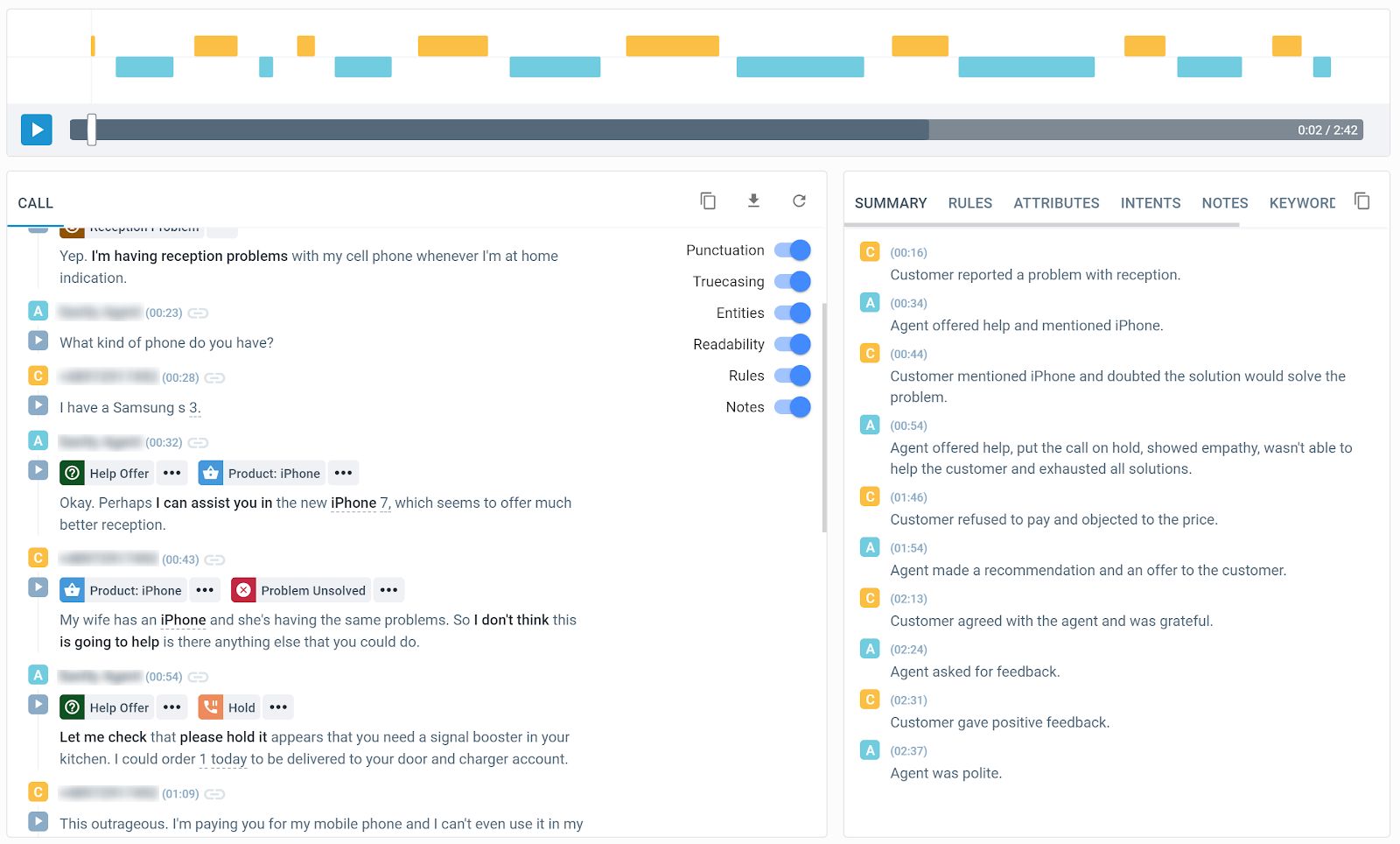}
\caption{Call Details panel with a fully transcribed, refined, and summarized call.}
\label{fig:call_details}
\end{figure}

\subsection{Streaming API}

Streaming API enables our customers to subscribe to the real-time stream of transcriptions, intents, entities, and other derivative events to build their own applications, such as agent assistants, robotic process automation, or custom live-supervision dashboards.



\bibliographystyle{IEEEtran}
\bibliography{mybib}

\end{document}